\documentclass[superscriptaddress,twocolumn,showpacs,preprintnumbers,amsmath,amssymb]{revtex4}

\usepackage{graphicx}

\newcommand{\etal}{\textit{et al.}}
\newcommand{\BABAR}{BABAR Collaboration}
\newcommand{\Belle}{Belle Collaboration}

\newcommand{\PRL}[3]{Phys. Rev. Lett. \textbf{#1}, (#2) #3}
\newcommand{\hepex}[1]{\eprint{hep-ex/#1}}

\begin{document}

\preprint{}

\title{
CP Asymmetry in $\bar B^0\to K^- \pi^+$ from SUSY Flavor Changing Interactions}

\author{Xiao-Gang He}
\email[e-mail: ]{hexg@phys.ntu.edu.tw}
\affiliation{Department of Physics, Peking University, Beijing 100871, China}
\affiliation{NCTS/IPE, Department of Physics, National Taiwan University, Taipei}
\author{Chong-Sheng Li}
\email[e-mail: ]{csli@pku.edu.cn}
\author{Li-Lin Yang}
\email[e-mail: ]{freewill@pku.edu.cn}
\affiliation{Department of Physics, Peking University, Beijing 100871, China}


\date{\today}

\begin{abstract}
Recently Babar and Belle collaborations have measured direct CP
asymmetry $A_{CP}=-0.114\pm 0.020$ in $\bar B^0 \to K^- \pi^+$.
The experimental value is substantially different from QCD
factorization prediction. We show that SUSY flavor changing
neutral current interaction via gluonic dipole can explain the
difference. CP asymmetries in other $B\to K\pi$ decays are
predicted to be sizeable. Taking this asymmetry as a constraint,
we find that the allowed SUSY parameter space is considerably
reduced compared with constraint from $B\to X_s \gamma$ alone.  We
also find that the allowed time dependent CP asymmetries $S_f$ in
$\bar B^0 \to \bar K^{* 0}\gamma \to \pi^0 K_S \gamma$ and $\bar
B^0\to \phi K_S$ to be large. These predictions are quite
different than those predicted in the Standard Model and can be
tested in the near future.
\end{abstract}

\maketitle

Recently Babar and Belle collaborations have measured direct CP
asymmetry $A_{CP}$ in $\bar B^0 \to K^- \pi^+$ with a value of
$-0.114\pm 0.020$\cite{expcpasy}. With also precision
determinations of the branching ratios of $B\to X_s\gamma$, $B\to
K\pi$, and other rare $B$ decays\cite{hfag,cpgamma,babar}, the
study of rare $B$ decays has entered a precision era. These decays
being rare in the Standard Model (SM) are very sensitive probes
for new physics beyond the SM.

The recently measured CP asymmetry $A_{CP}(\bar B^0 \to K^-
\pi^+)$ has important implications for $B$ decays and there have
been some discussions in the literatures\cite{implication}. The
experimental value for $A_{CP}(\bar B^0 \to K^- \pi^+)$ is
substantially different from the predictions based on
factorization calculations which predict $A_{CP}(\bar B^0\to K^-
\pi^+)$ to be positive for a set of favored hadronic parameters
\cite{bbns,du}. Although at this stage one
cannot conclude that there is the need of new physics beyond the
SM, as there are also methods which can give a value close to the
experimental data, such as pQCD calculations\cite{li}. We emphases
that at present there is not a method which can explain all data,
branching ratios and CP asymmetries in B decays simultaneously. It
is, nevertheless, important to see what new physics may be needed
to explain the data and what predictions can be made by
consistently using one method. When combined with other processes
which are more hadronic model independent crucial information
about new physics beyond SM can be extracted.

In this letter we take such an approach to study implications of
the CP asymmetry in $\bar B^0 \to K^- \pi^+$ on supersymmetric
(SUSY) flavor changing neutral current (FCNC) interaction via
gluonic dipole term using QCD improved factorization. We combine
more hadronic model independent process $B\to X_s\gamma$ to
constrain the relevant parameters. Predictions for direct CP
asymmetries in other $B\to K\pi$ decays, and time dependent CP
asymmetries in $B\to K^* \gamma \to \pi K_S\gamma$ and $B \to \phi
K_S$ are also studied.

In the SM, the Hamiltonian for the $B$ decays to be considered is
well known which is of the form\cite{sm},
\begin{eqnarray}
H={G_F\over \sqrt{2}} [
V_{ub} V^*_{us}(c_1 O_1 +c_2 O_2)-
\sum^{12}_{i=3}V_{jb}V_{js}^*c^j_iO_i],
\end{eqnarray}
where $V_{ij}$ are the CKM matrix elements. $c_i$ are the Wilson
coefficients for the operators $O_i$ which have been evaluated in
different schemes, values of which from NDR scheme will be
used\cite{sm}. We will not display the full sets of $O_i$ and
$c_i$ here, but only give the definitions of the gluonic and
photonic dipole operators $O_{11}$ and $O_{12}$ for the
convenience of later discussions. They are given by
\begin{eqnarray}
O_{11} &=& {g_s\over 8\pi^2} \bar s \sigma_{\mu\nu}
G^{\mu\nu}_a T^a[m_b(1+\gamma_5)+m_s(1-\gamma_5)]b,\nonumber\\
O_{12} &=& {e\over 8\pi^2} \bar s \sigma_{\mu\nu} F^{\mu\nu}
[m_b(1+\gamma_5)+m_s(1-\gamma_5)]b,
\end{eqnarray}
where $T^a$ is the color SU(3) generator normalized to $Tr(T^a
T^b) = \delta^{ab}/2$. $G_{\mu\nu}$ and $F_{\mu\nu}$ are the gluon
and photon field strengths. In the SM $c_{11}=-0.151$ and $c_{12}
= -0.318$\cite{sm,sm1,kagan}.

When going beyond the SM, there are some modifications of the
above coefficients. In SUSY models, exchanges of gluino and squark
with Left-Right squark mixing can generate a large contribution to
$c_{11,12}$ at one-loop level\cite{susydipole,massinsertion} since
their interactions are strong couplings in strength and also
enhanced by a factor of the ratio of gluino mass to the $b$ quark
mass\cite{pheno}. We will concentrate on the effects of this
interaction, although there are also possible large contributions
from other sources\cite{garisto}. In general exchange of squarks
and gluinos can generate non-zero $c_{11,12}$ for dipole operators
with $1+\gamma_5$, as well as with non-zero $c'_{11,12}$ for
dipole operators with $1-\gamma_5$.

The Wilson coefficients $c^{susy}_{11,12}$ from SUSY contributions
obtained in the mass insertion approximation are given by, for the
case with $1+\gamma_5$ \cite{massinsertion},
\begin{eqnarray}
&&c_{11}^{susy}(m_{\tilde g}) = {\sqrt{2}\pi \alpha_s(m_{\tilde
g})\over G_F m^2_{\tilde g}} {\delta_{LR}^{bs}\over
V_{tb}V_{ts}^*}{m_{\tilde g}\over m_b}
G_0(x_{gq}),\nonumber\\
&&c_{12}^{susy}(m_{\tilde g}) = {\sqrt{2}\pi \alpha_s(m_{\tilde
g})\over G_Fm^2_{\tilde g}} {\delta_{LR}^{bs}\over
V_{tb}V_{ts}^*}{m_{\tilde g}\over m_b}
F_0(x_{gq}),\nonumber\\
&&G_0(x) = {x[22-20x-2x^2+(16x-x^2+9)\ln(x)] \over 3(1-x)^4},
\nonumber\\
&&F_0(x) = -{4x[1+4x-5x^2+(4x +2x^2)\ln(x)] \over 9(1-x)^4},
\label{dipole}
\end{eqnarray}
where $\delta_{LR}^{bs}$ parameterizes  the mixing of left and
right squarks, $x_{gq} = m^2_{\tilde g}/m^2_{\tilde q}$ is the
ratio of gluino mass $m_{\tilde g}$ and squark mass $m_{\tilde
q}$. The Wilson coefficients $c^{'susy}_{11,12}$ for the case with
$1-\gamma_5$ can be easily obtained by replacing the Left-Right
mixing parameter $\delta_{LR}^{bs}$ by the Right-Left mixing
parameter $\delta^{bs}_{RL}$.

At the energy scale relevant for $B$ decays,
$\mu\approx m_b$, the coefficients $c^{(')susy}_{11,12}$ are
modified to be\cite{sm}, $c^{(')susy}_{11}(\mu)$ $=$ $\eta^7
c_{11}^{(')susy}(m_{\tilde g})$,
 and $c^{(')susy}_{12}(\mu)$ $=$ $\eta^8 c^{(')susy}_{12}(m_{\tilde
g})$$+$$ {8\over 3}(\eta^7 - \eta^8)c^{(')susy}_{11}(m_{\tilde
g})$, with $\eta$$ =$$ (\alpha_s(m_{\tilde
g})/\alpha_s(m_t))^{2/21} (\alpha_s(m_t)/\alpha_s(m_b))^{2/23}$.

From the expressions in Eq.(\ref{dipole}), one can see that the
SUSY contributions are proportional to $m_{\tilde g}$. If
$m_{\tilde g}$ is of order a few hundred GeV, there is an
enhancement factor of $(m_{\tilde g}/m_b) (m_W^2/m_{\tilde g}^2)$
for the SUSY dipole interactions. In this case even a small
$\delta_{LR, RL}^{bs}$, which can easily satisfy constraints from
$B^0-\bar B^0$ mixing and other data, can have large effects on
rare $B$ decays.

We first consider constraint on the SUSY parameters
$\delta^{bs}_{LR,RL}$ from $B\to X_S \gamma$. The branching ratio
of this process has been measured to a good precision with
$(3.54^{+0.30}_{-0.28})\times 10^{-4}$\cite{hfag}. Theoretically
the branching ratio has been evaluated to the next-to-leading
order QCD corrections. The branching ratio with the photon energy
cut to have $E_\gamma
> (1-\delta) E_\gamma^{max}$ is given by\cite{kagan}

\begin{eqnarray}
2.57\times 10^{-3}\times K_{NLO}(\delta)\,\times \,
  \frac{Br(B\rightarrow X_{c}e\overline{\nu })}{10.5\%},
\label{br}
\end{eqnarray}
 where the factor
$K_{NLO}(\delta)$ related to the Wilson coefficients $c_i$ is
given by, $K_{NLO}(\delta )=\sum_{\stackrel{i,j=2,11,12}{i\leq
j}}k_{ij}(\delta ) [c_{i}c_{j}^{*}+c_{i}^{\prime }c_{j}^{\prime
*}]+k_{12,12}^{(1)}(\delta )
[c_{12}^{(1)}c_{12}^{*}+c_{12}^{\prime (1)}c_{12}^{\prime *}]$.
The values of $c_{2}^{\prime }$ and $k_{ij}(\delta )$ can be
obtained by using the expressions given in Ref.~\cite{kagan}. We
use $\delta =90\%$ which gives $Br(B\to X_S \gamma) \approx
3.5\times 10^{-4}$, which is consistent with the data and the
complete NLO QCD results in Ref.~\cite{buras}.

Although experimentally CP asymmetry in $B\to X_S\gamma$ has not
been well established, there are constraints from experiments with
$0.005 \pm 0.036$ \cite{cpgamma}. We will also take this
information into account. In the SM, the leading contribution to
$A_{CP}(B \to X_S\gamma)$ is given by
\begin{eqnarray}
A_{X_S\gamma} &=&
  \frac{1}{|c^{SM}_{12}|^2}
  \{a_{27}Im[c^{SM}_2c^{SM*}_{12}]\nonumber\\
&+&a_{28}Im[c^{SM}_2c^{SM*}_{11}]
  +a_{87}Im[c^{SM}_{11}c^{SM*}_{12}]
\}. \label{gammacp}
\end{eqnarray}
From Ref.\cite{kagan}, we find $a_{87}\sim -9.5\%$, $a_{27}\sim
1.06\%$, and $a_{28} \sim 0.16\%$. For the calculation of
$A_{CP}(B\to X_S\gamma)$ in SUSY model considered here, one just
replaces $c^{SM}_{11,12}$ by the total $c_{11,12}$ and adds a term
$a_{87}Im[c'_{11}c'^{*}_{12}]$ to the numerator and $|c'_{12}|^2$
to the denominator in the above equation.

Using the above, deviations of $c^{(')}_{11,12}$ from the SM
values are severely constrained. In Figure 1 we show the allowed
ranges for the absolute values of $\delta^{bs}_{LR,RL}$ and their
phases $\tau$ for $m_{\tilde g}=300$GeV and $m_{\tilde q}$ in the
range $100 \sim 1000$GeV at the one $\sigma$ level. We find that
the constraints
from $Br(B\to X_S\gamma)$ are slightly more stringent that those from
$A_{CP}(B\to X_S\gamma)$. Using the allowed parameters, one can
obtain the allowed $c^{(')}_{11}$ through Eq.(\ref{dipole}) and to
study implications for other rare $B$ decays.
The allowed ranges
for $\delta^{bs}_{LR} (\tau_{LR})$ and
$\delta^{bs}_{RL} (\tau_{RL})$
are correlated in general.


\begin{figure}[htb]
\begin{center}
\includegraphics[width=4cm]{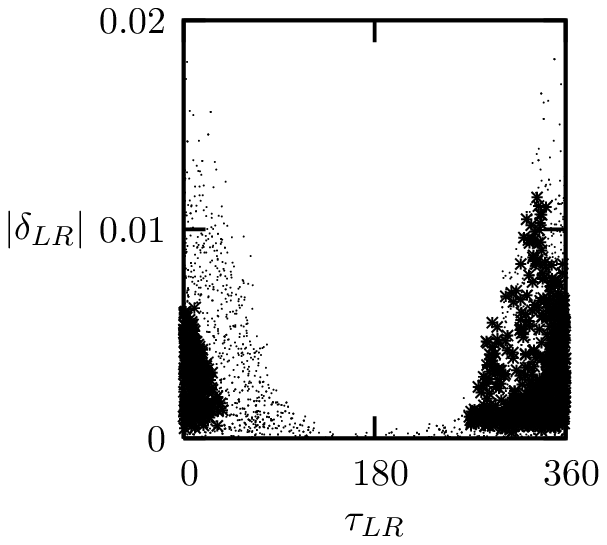}
\includegraphics[width=4cm]{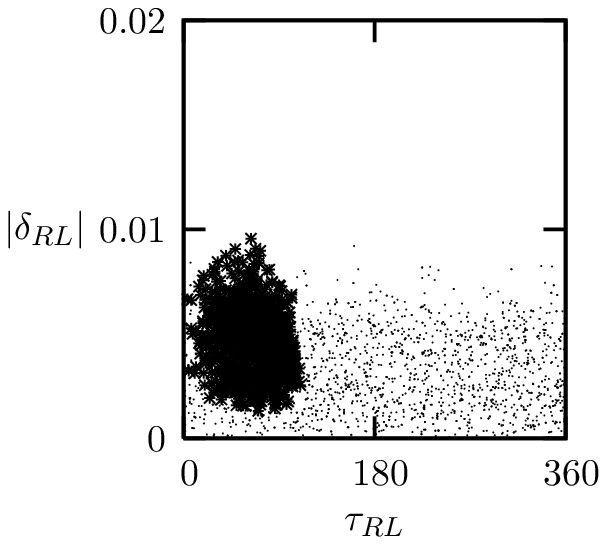}
\end{center}
\caption{The one $\sigma$ allowed ranges for the SUSY parameters
$|\delta_{LR,RL}^{bs}|$ and the phase $\tau$ taking $m_{\tilde
g}=300$GeV and $m_{\tilde q}$ in the range $100 \sim 1000$ GeV.
The light-dark doted areas are the allowed parameter spaces from
$Br(B\rightarrow \, X_s \gamma)$ and $A_{CP}(B\to X_S\gamma)$
constraints. The dark doted areas are allowed ranges by $A_{CP}(
\bar B^0 \to K^- \pi^+)$ constraint. Figure 1a (on the left)
 and Figure 1b (on the right) are for the
dipole operators with $1+\gamma_5$ and $1-\gamma_5$,
respectively.}
\end{figure}

We now study whether after the constraint from $B\to X_S \gamma$,
the observed CP asymmetry $A_{CP}(\bar B^0 \to K^- \pi^+)$ can be
reproduced and analysis what constraint can be put on the SUSY
parameters. For this purpose, we need to calculate the amplitudes
for $B\to K\pi$ decays. We follow the QCD factorization approach
in Ref.\cite{bbns}. Since $O_{12}$ is suppressed by a factor of
$\alpha_{em}/\alpha_s$ compared with $O_{11}$, we will neglect its
contribution in our later discussions. We find that the gluonic
dipole contributions to the decay amplitudes: $-\sqrt{2}A(\bar
B^0\to \bar K^0 \pi^0)$, $A(\bar B^0\to K^- \pi^+)$,
$\sqrt{2}A(B^-\to K^- \pi^0)$, $A(B^-\to \bar K^0 \pi^-)$ are the
same, which are given by

\begin{eqnarray}
i{G_F\over \sqrt{2}} f_K m^2_BF_0^{B\to \pi}(m^2_K) V_{tb}V_{ts}^*
{C_F\alpha_s\over 2\pi N_c}(c_{11}-c'_{11})G_{K\pi}, \label{kpi}
\end{eqnarray}
where $G_{K\pi}= \int^1_0 \phi_K(x)dx/(1-x) + R_K $. $R_K =
2m^2_K/m_s m_b$, and $C_F = (N_c^2-1)/(2N_c)$ with the number of
color  $N_c=3$. $\phi_K(x)$ is the light cone distribution
amplitude.

In our numerical analysis we will take the CKM parameters to be
known, with the standard parametrization $s_{12}=0.2243$,
$s_{23}=0.0413$, $s_{13}=0.0037$, $\delta_{13}=1.05$, which is the
central value given by the Particle Data Group\cite{pdg}. With the
SM amplitudes obtained and the default values for the hadronic
parameters used in Ref.~\cite{bbns}, we obtain the CP asymmetry
$A_{CP}(\bar B^0 \to K^- \pi^+)$ in the SM to be 0.15. This is
different in sign with the experimental value. When SUSY dipole
interactions are included the experimental value can be
reproduced. For example
$m_{\tilde{g}}=m_{\tilde{q}}=300$GeV, $\delta_{LR}=2.62\times
10^{-3}e^{0.238i}$, $\delta_{RL}=4.31\times 10^{-3}e^{1.007i}$ the
asymmetry $A_{CP}(\bar B^0 \to K^- \pi^+)$ is approximately
$-0.114$. Using the same set of SUSY parameters, we have $Br(B\to
X_s\gamma)=3.48\times 10^{-4}$, $A_{CP}(B\to X_S \gamma)=0.016$.
It is clear that the CP asymmetry $A_{CP}(\bar B^0\to K^- \pi^+)$
can be brought to be in agreement with data at one $\sigma$ level
when SUSY gluonic dipole interactions are included.

To see how the CP asymmetry provides stringent constraint on the
SUSY flavor changing parameters, we show in Figure 1 the parameter
space allowed from $A_{CP}(\bar B^0\to K^- \pi^+)$ (the dark doted
areas) on top of the allowed ranges by $B\to X_S \gamma$
constraint alone at the one $\sigma$ level. We see that the CP
asymmetry in $\bar B^0 \to K^- \pi^+$ considerably reduces the
allowed parameter space.

Using the above allowed SUSY parameters, one can predict the
branching ratios for all the four $B\to K\pi$ branching ratios and
also the unmeasured CP asymmetries. Since the branching ratios
involve unknown $B\to K$ and $B\to \pi$ form factors, one cannot
make precise predictions without a good understanding of these
form factors. We therefore study just the CP asymmetries in which
large part of the form factor effects are cancelled out. In Figure
2, we show the direct CP asymmetries in $B^- \to \bar K^0 \pi^-$,
$B^- \to K^- \pi^0$ and $\bar B^0\to \bar K^0 \pi^0$ for the
allowed parameter space in Figure 1. We see that large CP
asymmetries are allowed. In particular that the CP asymmetry in
$B^-\to \bar K^0 \pi^-$ can be as large as $-0.3$, whereas in the
SM this asymmetry is very small. Near future experiments can test
these predictions.


\begin{figure}[htb]
\begin{center}
\includegraphics[width=2.5cm]{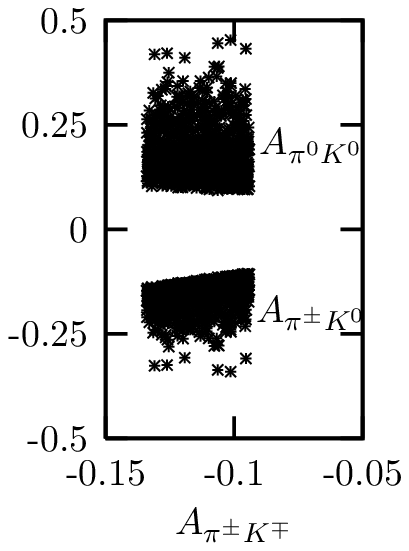}
\includegraphics[width=2.5cm]{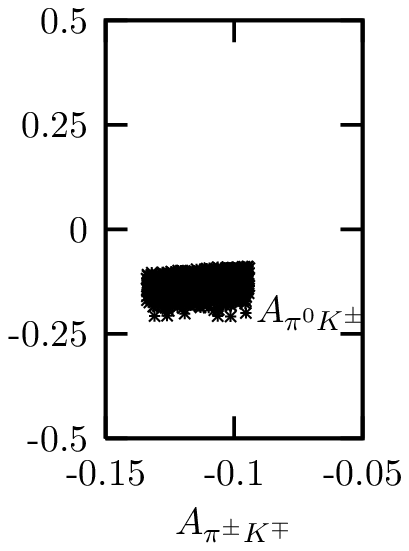}
\end{center}
\caption{The allowed CP asymmetries in $ B^- \to \bar K^0 \pi^-$,
$B^-\to K^- \pi0$ and $\bar B^0 \to \bar K^0 \pi^0$. }
\end{figure}

We finally study time dependent CP asymmetries in $B\to
K^*\gamma\to \pi^0 K_S \gamma$ and $B\to \phi K_S$. There are two
CP violating parameters $A_f$ and $S_f$ which can be measured in
time dependent decays of $B$ and $\bar B$ produced at $e^+e^-$
colliders at the $\Upsilon(4S)$ resonance, $A^{CP}(t) = A_f
\cos(\Delta t \Delta m_B) + S_f \sin(\Delta t \Delta m_B)$. The
parameters $A_f$ and $S_f$ are related to the decay amplitudes as
\begin{eqnarray}
A_f = \frac{|\lambda_f|^2 - 1}{|\lambda_f|^2 + 1},\;\;\;\;
S_f = -2\frac{Im[(q_B/p_B)\lambda_f]}{|\lambda_f|^2 + 1},
\end{eqnarray}
where $\lambda_f=\bar{A}/A$ and $\bar{A}$ and $A$ are the decay
amplitudes of $\bar{B}^0 \to f_{CP}$ and $B^0 \to f_{CP}$, respectively.
$q_B/p_B$ is the mixing parameter in $B-\bar B$ mixing.

For $\bar B^0 \to \bar K^{*0} \gamma \to \pi^0 K_S \gamma$ and
$B^0 \to K^{*0}\gamma \to \pi^0 K_S \gamma$, we
have\cite{kagan,pheno,atwood}
\begin{eqnarray}
S_{K^*\gamma} &=&
  -2\frac{Im[(q_B/p_B)(c_{12}c'_{12})]}{|c_{12}|^2+|c'_{12}|^2}.
\end{eqnarray}
To the leading order $A_{K^*\gamma}$ is the same as $A_{CP}(B\to
X_S \gamma)$. Note that the hadronic matrix element $<K^*|\bar s
\sigma^{\mu\nu} (1\pm \gamma_5) b|B>$ does not appear, which makes
the calculation simple and reliable. In order to have a non-zero
$S_{K^*\gamma}$ both $c_{12}$ and $c'_{12}$ cannot be zero.

In the SM the asymmetries $A_{K^*\gamma}$ and $S_{K^*\gamma}$ are
predicted to be small with $A^{SM}_{K^*\gamma} \approx 0.5\%$,
$S^{SM}_{K^*\gamma} \approx 3\%$\cite{kagan, atwood}. With SUSY
gluonic dipole interaction, the predictions for these CP
asymmetries can be changed dramatically\cite{pheno}. With the
constraints obtained previously, we find that the parameter
$q_B/p_B$ is not affected very much compared with the SM
calculation. To a good approximation $q_B/p_B =e^{-2i\beta}$.

A large gluonic dipole interaction also has a big impact on $B\to
\phi K_S$ decays\cite{phiK}. In the SM, $A_{\phi K_S}$ is
predicted to be very small and $S_{\phi K_S}$ is predicted to be
the same as $S_{J/\psi K_S}=\sin (2\beta)$. With SUSY gluonic
dipole contribution, the decay amplitude for $B\to \phi K_S$ will
be changed and the predicted value for both $A_{\phi K_S}$ and
$S_{\phi K_S}$ can be very different from those in the
SM\cite{phiK}. To obtain concrete values, we again use QCD
factorization to evaluate the amplitude. We obtain the
contributions of $c_{11}$ and $c'_{11}$ to $B\to \phi K_s$
amplitude to be
\begin{eqnarray}
{G_F\over \sqrt{2}} m_\phi f_\phi F^{B\to K}_1(m^2_\phi)
\epsilon^\mu_\phi \cdot (P^B_\mu + P^K_\mu) (c_{11} + c'_{11})
G_{\phi,11},
\end{eqnarray}
where $\epsilon^\mu_\phi$ is the polarization vector of $\phi$.
$G_{\phi,11} = -\int^1_02\phi_\phi(x)dx/(1-x)$ with $\phi_\phi(x)$
being the light cone distribution function.


\begin{figure}[htb]
\begin{center}
\includegraphics[width=4cm]{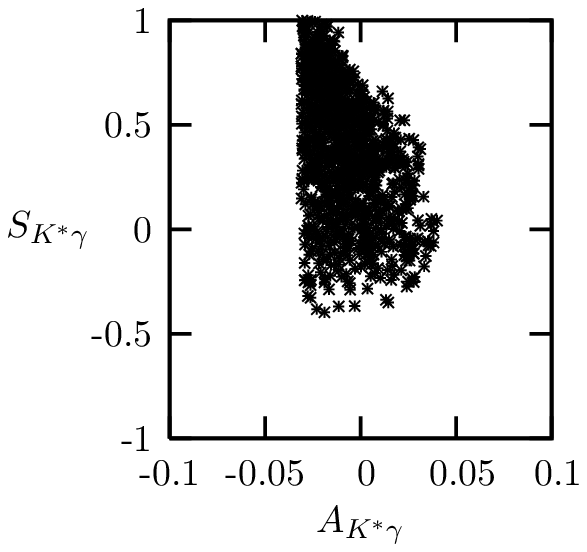}
\includegraphics[width=4cm]{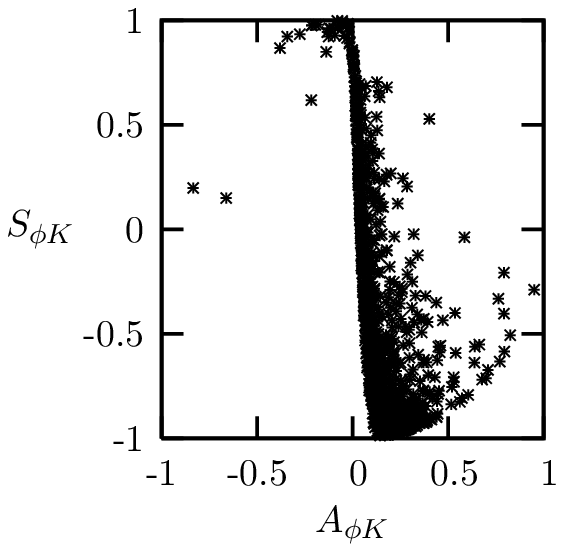}
\end{center}
\caption{The allowed time dependent CP asymmetries in $ \bar B^0 \to K^*\gamma\to
K_S \pi^0 \gamma$ and $\bar B^0\to \phi K_S$. }
\end{figure}

We are now ready to present the allowed ranges for the time
dependent parameters $A_f$ and $S_f$ for both the processes $\bar
B^0 \to \bar K^* \gamma \to K_S \pi^0 \gamma$ and $\bar B^0 \to
\phi K_S$. The results are shown in Figure 3. The current values
of $S_{K^*\gamma}$ and $A_{K^*\gamma}$ from Babar (Belle) are:
$0.57\pm 0.32\pm 009(-0.00\pm 0.38)$, $0.25\pm
0.63\pm0.14(-0.79^{+0.63}_{-0.50}\pm 0.09)$, respectively
\cite{kgamma}. From Figure 3, we see that the allowed ranges can
cover the central values of $S_{K^*\gamma}$ from Babar and Bell,
but it is not possible to obtain the central value of
$A_{K^*\gamma}$ by Belle. Future improved data can further
restrict the parameter space. Both Babar and Belle have also
measured $A_{CP}(B^- \to K^{*-}\gamma)$ with ranges $-0.074 \sim
0.049$(Babar) and $-0.015\pm 0.044\pm 0.012$(Belle) \cite{ss}. In
the model we are considering, the CP asymmetries $A_{K^*\gamma}$
and $A_{CP}(B^- \to K^{*-}\gamma)$ are the same. The results for
the charged $B$ CP asymmetry are consistent with data.

The time dependent asymmetry in $B\to \phi K_S$ is a very good
test of CP violation in the SM. Experimental measurements have not
converged with the current values of Babar(Belle) given by
$0.00\pm0.23\pm0.05(0.08\pm 0.22\pm 0.09)$, and
$0.50\pm0.25^{+0.07}_{-0.04}(0.06\pm0.33\pm0.09)$ for $A_{\phi
K_S}$ and $S_{\phi K_S}$\cite{kphi,babar}, respectively. These
values are considerably different than the value reported by Belle
last year of $S_{\phi K_S} = -0.96\pm
0.50^{+0.09}_{-0.11}$\cite{belle03}. From Figure 3 we see that the
current data of $A_{\phi K_S}$ and $S_{\phi K_S}$ can be easily
accommodated by the allowed ranges. We also note that the allowed
ranges can cover last year's Belle data. Since the error bars on
the data are large, no definitive conclusions can be drawn at
present.


In summary we have studied the implications of the recently
measured CP asymmetry in $\bar B^0\to K^- \pi^+$ on SUSY flavour
changing interactions. The experimental value for this asymmetry
$-0.114\pm 0.020$ is substantially different than QCD
factorization prediction. We have shown that SUSY FCNC interaction
via gluonic dipole can explain this difference. The allowed SUSY
parameter space is considerably reduced compared with constraint
from $B\to X_s \gamma$ alone. CP asymmetries in other $B\to K\pi$
decays are predicted to be sizeable. We also find that the allowed
time dependent CP asymmetries $S_f$ in $\bar B^0 \to \bar K^{*
0}\gamma \to \pi^0 K_S \gamma$ and $\bar B^0\to \phi K_S$ are in
the ranges of $-0.4 \sim 1$ and $-1 \sim 1$, respectively. These
predictions are quite different from the ones in the SM and can be
tested in the near future.

\begin{acknowledgments}
This work was supported in part by the National Natural Science
Foundation of China and Specialized Research Fund for the Doctoral
Program of Higher Education.
\end{acknowledgments}

\end{document}